\documentclass[12pt]{iopart}
\usepackage{graphicx}
\usepackage{hyperref}

\begin{document}

\title[ELM for Anomalous Diffusion]{Extreme Learning Machine for the Characterization of Anomalous Diffusion from Single Trajectories (AnDi-ELM)}

\author{Carlo Manzo}
\address{Facultat de Ci\`encies i Tecnologia, Universitat de Vic -- Universitat Central de Catalunya (UVic-UCC), C. de la Laura,13, 08500 Vic, Spain}
\ead{carlo.manzo@uvic.cat}
\vspace{10pt}

\begin{abstract}
The study of the dynamics of natural and artificial systems has provided several examples of deviations from Brownian behavior, generally defined as anomalous diffusion. The investigation of these dynamics can provide a better understanding of diffusing objects and their surrounding media, but a quantitative characterization from individual trajectories is often challenging. Efforts devoted to improving anomalous diffusion detection using classical statistics and machine learning have produced several new methods. Recently, the anomalous diffusion challenge (AnDi, \href{https://www.andi-challenge.org}{www.andi-challenge.org}) was launched to objectively assess these approaches on a common dataset, focusing on three aspects of anomalous diffusion: the inference of the anomalous diffusion exponent; the classification of the diffusion model; and the segmentation of trajectories.  In this article, I describe a simple approach to tackle the tasks of the AnDi challenge by combining extreme learning machine and feature engineering (AnDi-ELM). The method reaches satisfactory performance while offering a straightforward implementation and fast training time with limited computing resources, making it a suitable tool for fast preliminary screening of anomalous diffusion.

\end{abstract}

\bigskip
\noindent\textbf{Comments:} 
This is the Accepted Manuscript version of an article accepted for publication in 
\textit{Journal of Physics A: Mathematical and Theoretical}. 
IOP Publishing Ltd is not responsible for any errors or omissions in this version of the manuscript 
or any version derived from it. 
The Version of Record is available online at 
\href{https://doi.org/10.1088/1751-8121/ac13dd}{https://doi.org/10.1088/1751-8121/ac13dd}.

\vspace{2pc}
\maketitle
%
%

\section{Introduction}
From biology to the stock market, a variety of systems display dynamics that deviate from Brownian diffusion in the form of a nonlinear scaling of the mean squared displacement vs time, typically referred to as anomalous diffusion~\cite{metzler2014anomalous}. The detailed characterization of these processes can lead to important findings about the system components, their interactions, and the surrounding environment.

The development of experimental techniques with single-molecule detection capability has allowed tracing the motion of nanometric particles in living cells~\cite{manzo2015review}, often revealing the occurrence of anomalous diffusion~\cite{barkai2012single}. However, the analysis of these experiments can be rather challenging. In several cases, technical and instrumental limitations impact data collection and quality,  producing short and noisy trajectories that prevent a precise assessment of the underlying dynamics. In addition, the heterogeneous nature of the diffusing environment might further introduce subtle spatiotemporal changes that are difficult to isolate from noise~\cite{manzo2015weak,weron2017ergodicity, sabri2020elucidating}.

To limit these drawbacks, the development of improved methods for characterizing diffusion from single trajectories has proliferated, becoming a central topic in biophysics. New statistical approaches to assess anomalous diffusion have been proposed ~\cite{kepten2015guidelines,burnecki2015estimating, weron2017ergodicity, sikora2017elucidating, thapa2018bayesian, krapf2019spectral, thapa2020leveraging}. More recently, the popularization of machine learning (ML) has further stimulated contributions to this field through the development of data-driven algorithms~\cite{munoz2020single, granik2019single, bo2019measurement, kowalek2019classification,janczura2020classification,vangara2020identification}.    

The large number of contributions in this field has led to the organization of the Anomalous Diffusion challenge (AnDi, \href{https://www.andi-challenge.org}{www.andi-challenge.org})~\cite{munoz2020anomalous}, a competitive effort to foster the development of new approaches and assess their performance for the quantification of different aspects of anomalous diffusion on a common dataset. The challenge was divided into three tasks: the inference of the anomalous diffusion exponent; the classification of the underlying diffusion model; and the segmentation of trajectories presenting a change of anomalous diffusion exponent and/or model. Each task was further divided into subtasks corresponding to the trajectory dimensions (1D, 2D, and 3D).

As an organizer of the challenge, I had a great interest in the comparison of methods' performance, even beyond competition metrics. Specifically, I wondered whether it would be possible to obtain a respectable performance using an off-the-shelf ML approach that could be easily implemented by non-experts using limited computational resources. To this aim, I searched for a method complying with a list of requirements:  ML-based; easy to code and fine-tune; short computational time for training on a regular CPU; a limited number of hyperparameters; possibility to be applied to all the tasks of the challenge with minimal modifications. In addition, to allow a fair comparison, training must be performed using only the public datasets provided to challenge participants. 

The choice fell on Extreme Learning Machine (ELM), a learning algorithm for single-hidden layer feedforward networks (SLFNs) proposed by Huang and coworkers~\cite{huang2004extreme, huang2006extreme}.  In ELM, the output weights of hidden nodes are learned in a single step, which is essentially equivalent to learning a linear model. Therefore, ELM offers the advantage of low computational cost as compared to other ML approaches. Although its shallow architecture affects its performance for complex tasks, ELM has been applied to several real-world problems, including the identification of COVID-19 using features extracted from chest X-ray images
~\cite{rajpal2020cov}, showing a good generalization capability~\cite{huang2006extreme}. 

In this article, I describe AnDi-ELM, a method that combines ELM with feature engineering to tackle the 3 tasks of the AnDi challenge. Features were calculated using classical statistical estimators. For tasks 1 and 2, AnDi-ELM uses estimations of the scaling exponents of the moments of the distribution of displacements~\cite{ferrari2001strongly},  and the time-correlation of displacements; for task 3, the cumulative sum of squared displacements. Suitable architectures were chosen by estimating the generalization ability of several models, varying the number of hidden units and the size of the training dataset. When applied to the challenge dataset, AnDi-ELM showed respectable results, with a  MAE$=0.24$~--~$0.28$ for the regression of the anomalous diffusion exponent; a $F_1$-score$=0.74$~--~$0.85$ for model classification; and an overall RMSE$=57$~--~$60$ for changepoint detection. Best results were generally obtained for larger dimensions. The method is robust with respect to the magnitude of noise. When compared to other methods taking part in the challenge, AnDi-ELM ranked better for the classification task than the regression one. In the latter case, its score was strongly penalized by limitations in determining the anomalous diffusion exponent of some nonergodic models, such as and the annealed transient time motion and the superdiffusive scaled Brownian motion.
\section{Methods \label{sec:methods}}
\subsection{Datasets, preprocessing, and feature engineering}
Datasets included in this study correspond to simulated trajectories generated with the Python package {\tt andi-dataset}~\cite{munoz2020dataset} developed for the AnDi challenge. Training datasets were composed of $5000$, $10000$, and $50000$ trajectories for each task and dimension. Validation and test datasets consisted of $10000$ trajectories for each task and dimension. Each trajectory included walker's spatial coordinates ${\bf x}(t_i)$ sampled at different times $t_i$ with a unitary time interval $\Delta t$. Trajectories were generated according to the 5 theoretical models included in the AnDi challenge: continuous-time random walk (CTRW)~\cite{scher1975anomalous}, fractional Brownian motion (FBM)~\cite{mandelbrot1968fractional}, L\'{e}vy walk (LW)~\cite{klafter1994levy}, annealed transient time motion (ATTM)~\cite{massignan2014nonergodic}, and scaled Brownian motion (SBM)~\cite{lim2002self}. Overall, the anomalous diffusion exponent had values in the range $\alpha \in \left[0.05, 2\right]$ with steps of $0.05$, but it was limited to smaller ranges for specific models~\cite{munoz2020anomalous}. Trajectories were corrupted with additive Gaussian noise with zero mean and different values standard deviation ($0.1, 0.5$ or $1.0$).  For trajectories with standardized displacements, this provided signal-to-noise ratios ${\rm SNR} = 10,2$ or $1$.
Datasets for task 1 and 2 contained trajectories  of variable duration $L \in \left[10,1000\right]$. Datasets for task 3 contained trajectories of fixed duration $L = 200$ obtained by joining segments of two trajectories generated from different theoretical models and/or anomalous diffusion exponents. 

Simulated trajectories were first standardized to have a unitary standard deviation of displacements for $t_{\rm lag}=1$ and their starting coordinates were set to zero. Then, according to the specific task, different sets of features were calculated. For task 1 and 2, two estimators were calculated for different time-lags $t_{\rm lag}= k \Delta t$ with $k = 1,...,7$, corresponding to
\begin{equation*}
F_d\left( k \right) = \frac{ {\rm log}\left< | {\bf x}(t_i+k \Delta t) -{\bf x}(t_i) |^d \right> } { {\rm log}\left[(k +1)\Delta t \right]}
\end{equation*}
for $d=1,2$. The features $F_d\left( k \right)$ represent estimations of the scaling exponents of the first ($d=1$) and second moment ($d=2$) of the distribution of displacements obtained for different time-lags~\cite{ferrari2001strongly}.

In addition, the correlation of absolute displacements obtained for $k=1$ 
\begin{equation*}
G = \frac{ \left< | {\bf x}(t_i+ \Delta t) -{\bf x}(t_i) |\cdot | {\bf x}(t_j+ \Delta t) -{\bf x}(t_j) |  \right> } {\left< | {\bf x}(t_i+ \Delta t) -{\bf x}(t_i) |^2 \right>}
\end{equation*}
and with $t_j=t_i+\Delta t$ was further calculated, providing a total of 15 features per trajectory. Each feature was independently standardized using the $z$-score over the training dataset. The mean and standard deviation obtained for each feature of the training dataset were saved and later used to standardize the validation and test dataset. 

For task 3, features were calculated as the cumulative sum of squared displacements at all sampling times:
\begin{equation*}
D\left(t_i\right) =  \sum_{t_j=0}^{t_i}   | {\bf x}(t_j+k \Delta t) -{\bf x}(t_j) |^2
\end{equation*}
and they were rescaled to the range $[-1,1]$ for each trajectory. 

\subsection{ELM architecture}
\begin{figure}[!h]
    \centering
    \includegraphics[width=\textwidth]{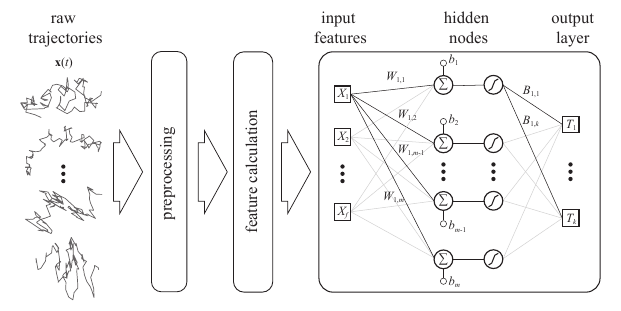}
    \caption{Schematic of the architecture of AnDi-ELM. Raw trajectory coordinates ${\bf x}(t)$ are standardized and used to calculate a set of $f$ features ${\bf X}$ for each trajectory. Features constitute the input of a SLFN with $m$ hidden nodes, with randomly initialized weights ${\bf W}$ and biases ${\bf b}$, and a sigmoid activation function. Elements of the output layer are obtained through the application of output weights ${\bf B}$. A common structure was used for all the tasks of the AnDi challenge but with different features, different number of hidden nodes and outputs.}
    \label{fig:scheme}
\end{figure}

For a training dataset of $n$ trajectories and $f$ features with $k$ target values ${\bf T}$, the $n \times f$ feature matrix ${\bf X}$ is fed into a SLFN with $m$ hidden nodes~\cite{huang2004extreme, huang2006extreme}(Fig.~\ref{fig:scheme}). A  matrix of initial weights ${\bf W}$ of size $f \times m $ and a bias vector ${\bf b}$ of size $1 \times m $ is randomly initialized to connect observations to targets through:
\begin{equation*}
    f\left( {\bf X} {\bf W} + {\bf u} {\bf b^{T}}\right) {\bf B}  = {\bf H} {\bf B} = {\bf T},
\end{equation*}
where ${\bf u}$ is a unitary vector of size $n \times 1$,  {\bf B} is the matrix of output weight, and $f\left( \cdot \right)$ represents the sigmoid activation function:
\begin{equation*}
f\left( x \right) = \frac{1}{1+e^{-x}}.
\end{equation*}
The training of the SFLN is thus converted into solving an over-determined linear problem, whose least-squares solution corresponds to the Moore-Penrose pseudoinverse of the hidden layer matrix {\bf H}~\cite{huang2004extreme, huang2006extreme}
\begin{equation*}
    {\hat{ \bf B }}={\bf H^{\dagger}}{\bf T}.
\end{equation*}

The SFLN was trained either as a regressor (task 1) or as a classifier (tasks 2 and 3) to provide predictions for all the tasks of the AnDi challenge. The target for task 1 simply consisted of a vector including the $\alpha$ value of each trajectory in the dataset. For tasks 2 and 3, targets were obtained through the one-hot encoding of the categorical variable, the latter having levels corresponding to the 5 diffusion models (task 2) or the position at which the changepoint occurred (task 3). 

The AnDi-ELM code was implemented using MATLAB R2020a (The MathWorks, Inc., Natick, Massachusetts, United States). The training was performed on a MacBookPro with an 8-Core Intel Core i9 processor with 2.4 GHz speed. Computational time is reported in Fig.~\ref{fig:hyperpar} as a function of the number of hidden layers $m$. The basic functions of the AnDi-ELM code are available online (\href{https://github.com/qubilab/AnDi_ELM}{github.com/qubilab/AnDi\_ELM}).

\subsection{Performance evaluation}
Different metrics were used to assess the performance of AnDi-ELM for the different tasks, according to those evaluated for the AnDi challenge. 
Results obtained for task 1 (inference of the anomalous diffusion exponent) were quantified by the mean absolute error (${\rm MAE}$) between the predicted value and the ground truth of the anomalous diffusion exponent:
\begin{equation}
	\label{eq:mae}
	{\rm MAE} = \frac{1}{N} \sum_{i=1}^{N}{| \alpha_{i,\rm{P}} - \alpha_{i,\rm{GT}} |},
\end{equation}
where $N$ is the number of trajectories in the dataset, $\alpha_{i,\rm{P}}$ and $\alpha_{i,\rm{GT}}$ represent the predicted and ground truth values of the anomalous exponent, respectively. 

For task 2 (diffusion model classification), I used the micro-averaged expression of the $F_1$-score:
\begin{equation}
	\label{eq:f1}
    F_1 = \frac{2\mbox{TP}}{2\mbox{TP}+\mbox{FP}+\mbox{FN}},
\end{equation}
where TP corresponds to the number of true positives, FP to false positives, and FN to false negatives.

Finally, for task 3 (trajectory segmentation) I assessed the precision for the location of the changepoint through the root mean squared error (RMSE) between the predicted and ground truth position:
\begin{equation} 
	\label{eq:rmse}
	{\rm RMSE} = \sqrt{\frac{1}{N} \sum_{i=1}^{N} \Big(t_{i,{\rm P}} - t_{i,{\rm GT}} \Big)^2},
\end{equation}
where $t_{i,{\rm P}}$ and $t_{i,{\rm GT}}$ represent the predicted and ground truth values of the changepoint  position, respectively.

\section{Results and Discussion}

\subsection{Determination of the optimal architecture}

\begin{figure}[!h]
    \centering
    \includegraphics[width=\textwidth]{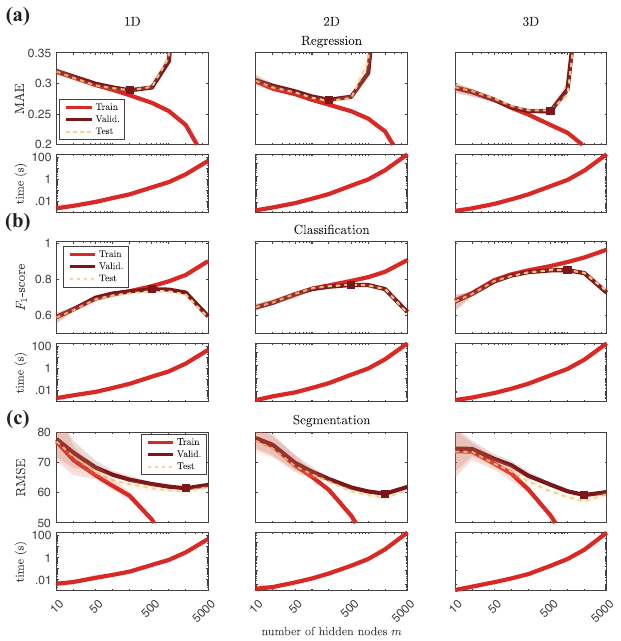}
    \caption{Determination of the optimal architecture. {\bf (a-c)} Semilog plots of performance metrics at varying the number of hidden nodes obtained by AnDi-ELM for training, validation, and testing datasets in 1D (left panels), 2D (central panels), and 3D (right panels). Lines correspond to average metrics over 10 runs. The shaded region embeds the full range of values. The training time as a function of the number of hidden nodes is also reported.  Symbols correspond to the optimal value obtained for the validation dataset. Metrics correspond to those used for performance assessment of the three tasks of the AnDi challenge:  MAE for regression {\bf (a)}, $F_1$-score for classification {\bf (b)}, and RMSE for segmentation {\bf (c)}.}
    \label{fig:hyperpar}
\end{figure}

I first sought to determine the optimal architecture of the ELM for each of the tasks/dimensions. In fact, a network with few nodes may not provide proper modeling of the data, whereas using too many nodes can produce overfitting. The low computational cost and the limited number of hyperparameters of ELM allow performing this evaluation by simply estimating the generalization capability for several models having a different number of hidden nodes. The AnDi-ELM was trained at varying the number of hidden nodes $m$ and its performance was evaluated by calculating the metrics corresponding to each task/dimension. Each training procedure was performed $10$ times to assess the effect of the random initialization of weights. Models were trained on a dataset composed of $10000$ trajectories and thus used to provide predictions for a validation and a test datasets. Training and validation datasets of $10000$ trajectories were generated using the Python package {\tt andi-dataset}~\cite{munoz2020dataset}. The test dataset was provided by the organizers of the AnDi challenge and ground truth values were made available after the conclusion of the competition. The results are shown in Fig.~\ref{fig:hyperpar}. 

As expected, the metrics for the training procedure show a monotonic trend as the number of hidden nodes is increased. The metrics obtained for the validation dataset follow a similar behavior for low $m$, to then invert their trend and diverge from the training curve, due to overfitting. I thus took the number of nodes corresponding to the meaningful extremum of the metrics (i.e., the minimum for MAE and RMSE, the maximum of $F_1$-score) as the optimal value of nodes of the Andi-ELM for each specific task/dimension. As shown in Fig.~\ref{fig:hyperpar},  the curves obtained for the test dataset follow a very similar trend as those obtained for the validation dataset. Moreover, the training time was below $100$ s even for the model with the largest number of hidden nodes ($m=5000$). For the regression and classification problems (Fig.~\ref{fig:hyperpar}(a-b)), the optimal $m$ value is very close to the onset of overfitting. However, AnDi-ELM shows larger overfitting for the segmentation problem. Although marginally, the RMSE of the validation dataset continues to decrease well after the learning curves start to diverge (Fig.~\ref{fig:hyperpar}(c)).

\begin{figure}[!h]
    \centering
    \includegraphics[width=\textwidth]{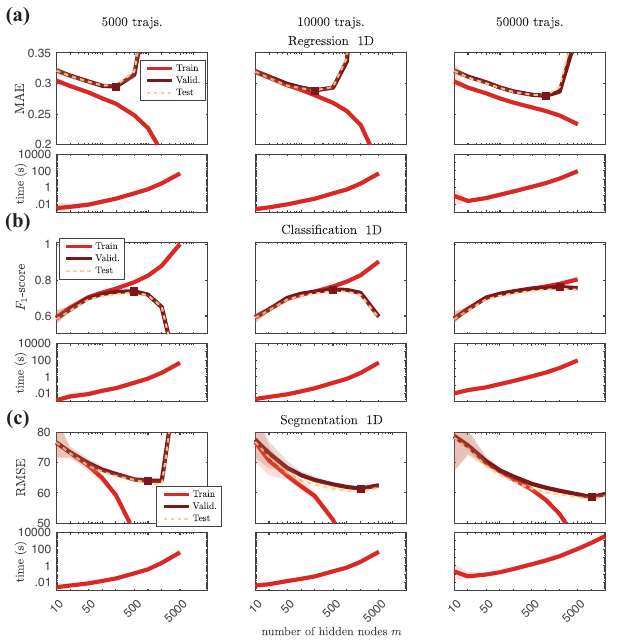}
    \caption{Effect of the size of the training dataset. {\bf (a-c)} Semilog plots of performance metrics at varying the number of hidden nodes obtained by AnDi-ELM for training, validation, and testing datasets in 1D using three different training datasets composed of $5000$ (left panels), $10000$ (central panels), and $50000$ trajectories (right panels). Lines correspond to average metrics over 10 runs. The shaded region embeds the full range of values. The training time as a function of the number of hidden nodes is also reported.  Symbols correspond to the optimal value obtained for the validation dataset. Metrics correspond to those used for performance assessment of the three tasks of the AnDi challenge:  MAE for regression {\bf (a)}, $F_1$-score for classification {\bf (b)}, and RMSE for segmentation {\bf (c)}.}
    \label{fig:hyperpar2}
\end{figure}

Furthermore, I also investigated the effect that the size of the training dataset has on the generalization capability and the performance of the methods. To this aim, two other training datasets composed of $5000$ and $50000$ trajectories were generated for the three tasks in 1D and used to train the AnDi-ELM at varying the number of hidden nodes $m$. The comparison of the learning curves in Fig.~\ref{fig:hyperpar2} shows that larger training datasets enable models with a higher number of hidden nodes and improve the generalization capability of the model since training and validation curves start diverging at larger values of $m$.  However, larger training datasets only produce a minor improvement in the method performance. On the other hand, architectures with a higher number of hidden nodes require larger computational time for training. For example, a five-fold increase from $10000$ to $50000$ trajectories allows a relative improvement of metric values below $5$\% ($3.1$\% for the regression, $1.8$\% for the classification, and $4.6$\% for the segmentation task in 1D). This minor improvement comes at the cost of a reduction of speed of about two orders of magnitude. In the case of the segmentation task, the training time goes from $3$ s to $10$ min. Based on these results, the rest of the study was carried out using the architecture and the model determined with the training dataset composed of $10000$ trajectories. 

\subsection{Inference of the anomalous diffusion exponent}

\begin{figure}[!h]
    \centering
    \includegraphics[width=1\textwidth]{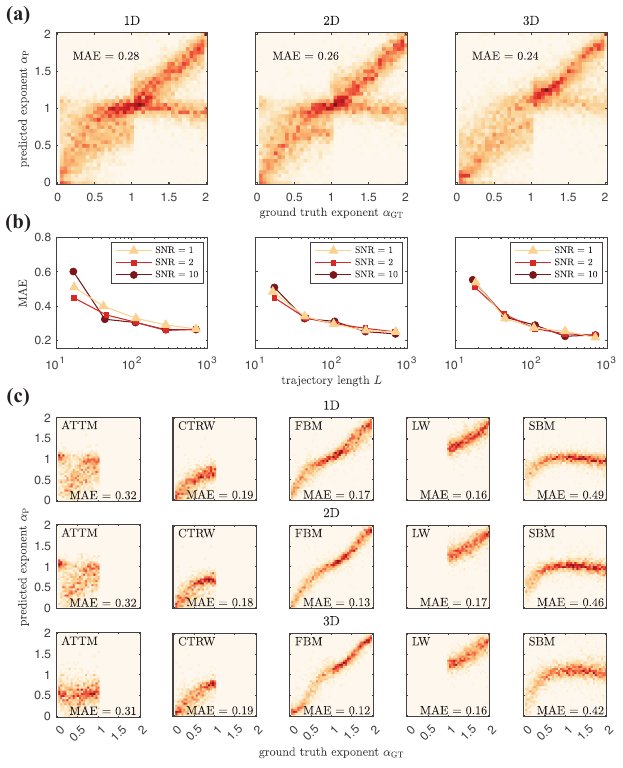}
    \caption{Results of the application of AnDi-ELM for the inference of the anomalous diffusion exponent. {\bf (a)} Two-dimensional histograms of the frequency of predicted anomalous diffusion exponent $\alpha_{\rm P}$ as a function of the ground-truth value $\alpha_{\rm GT}$ for trajectories of walkers undergoing anomalous diffusion in 1, 2, and 3 dimensions. {\bf (b)} Mean-absolute error (MAE) between $\alpha_{\rm P}$ and $\alpha_{\rm GT}$ as a function of trajectory length and for different levels of noise in 1, 2, and 3 dimensions. {\bf (c)} Two-dimensional histograms of the frequency of predicted anomalous diffusion exponent $\alpha_{\rm P}$ as a function of the ground-truth value $\alpha_{\rm GT}$ as in {\bf (a)} for each of the 5 diffusion models in 1, 2, and 3 dimensions.}
    \label{fig:inference}
\end{figure}

I thus focused on the analysis of the predictions provided by the optimal AnDi-ELM models for the inference of the anomalous diffusion exponent. The overall results are shown in the 2-dimensional histograms of predicted versus ground-truth anomalous diffusion exponent of Fig.~\ref{fig:inference}(a). 

In general, predicted values show a good level of correlation with the ground truth, with major deviations occurring for $1 < \alpha_{\rm GT} < 2$, where a fraction of superdiffusive trajectories are predicted as nearly Brownian ($\alpha_{\rm P} \sim 1$. Similar results were obtained for each problem dimensionality. A slight improvement in the MAE is observed as the number of dimensions is increased, likely as a consequence of the larger amount of information contained in multidimensional trajectories. 

Next, I sought to investigate the dependence of the AnDi-ELM performance as a function of trajectory length and SNR (Fig.~\ref{fig:inference}(b)). AnDi-ELM accuracy does not show any major dependence on the noise used to corrupt trajectories. However, performance markedly depends on trajectory length, with a two-fold decrease of MAE as trajectory length increases over two orders of magnitude. 

Last, as shown in Fig.~\ref{fig:inference}(c), I dissected the results as a function of the diffusion model used to simulate the trajectories. Interestingly, AnDi-ELM shows good performance (MAE$<0.2$) for CTRW, FBM, and LW trajectories. In the three cases, predicted values are correlated with the ground truth. For CTRW and LW, whose $\alpha_{\rm GT}$ values are limited to the sub- or super-diffusive range, respectively, predictions either underestimate or overestimate the true values for $\alpha_{\rm GT} \sim 1$. This is not the case for FBM, for which $\alpha_{\rm P}$ correlates with $\alpha_{\rm GT}$ over the full range of exponent $(0,2)$,  with a slight overestimation only observed in the subdiffusive range. In contrast, the prediction accuracy of the AnDi-ELM is rather poor for trajectory performing ATTM and SBM. For ATTM, predicted exponent are scattered over a wide range of values. Strongly subdiffusive trajectories are interpreted as Brownian motion. A different pattern is observed in 3D in contrast to 1D and 2D, where the method tends to return $\alpha_{\rm P} \sim 0.5$, independently of the ground-truth value.  For subdiffusive SBM, the method produces correlated although overestimated predictions. However, AnDi-ELM is not able to detect the superdiffusive behavior for this diffusion model, returning $\alpha_{\rm P}$ values compatible with Brownian diffusion for $\alpha_{\rm GT} > 1$. 

A major determinant for the correct evaluation of the anomalous diffusion exponent appears to be the ergodicity of the underlying model. This result can be at least in part ascribed to the choice of the input features $F_d\left( k \right) $,  which are strictly linked to the scaling exponents of the moments of the distribution of displacements. For ergodic models, the scaling exponent of the time-averaged mean squared displacement corresponds to the anomalous diffusion exponent. In contrast, weakly non-ergodic models such as SBM, ATTM, and CTRW, display a time-averaged mean squared displacement with a linear behavior~\cite{jeon2014scaled,manzo2015weak}. Based on the exponent of the time-averaged mean squared displacement, trajectories generated using weakly non-ergodic models might thus be erroneously identified as Brownian motion. Nonetheless, for CTRW, subdiffusive FBM, and partly for ATTM, the method can learn patterns of trajectory features that produce a reliable estimation of the anomalous diffusion exponent. Yet, the best performance is achieved for the ergodic FBM.

\subsection{Classification of the diffusion model}

\begin{figure}[h!]
    \centering
    \includegraphics[width=\textwidth]{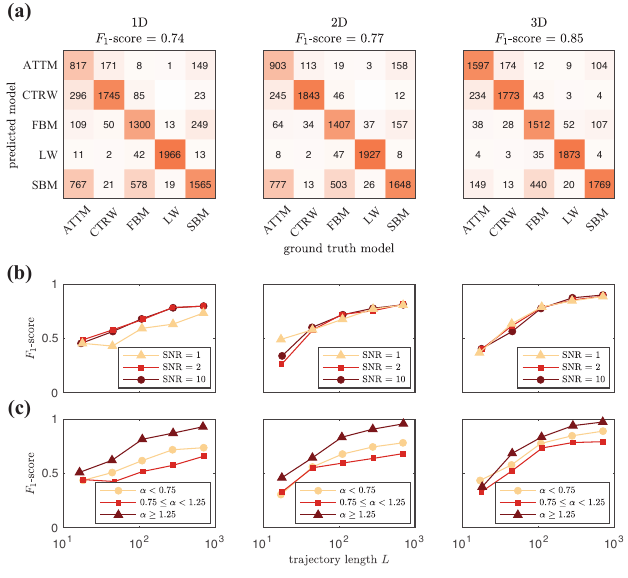}
    \caption{Results of the application of AnDi-ELM for the classification of the anomalous diffusion model. {\bf (a)} Confusion matrices of the number of predicted anomalous diffusion models as a function of the ground-truth model for trajectories of walkers undergoing anomalous diffusion in 1, 2, and 3 dimensions. {\bf (b)} $F_1$-score for model classification as a function of trajectory length and for different levels of noise in 1, 2, and 3 dimensions. {\bf (c)} $F_1$-score for model classification as a function of the range of anomalous diffusion exponent $\alpha$ in 1, 2, and 3 dimensions.}
    \label{fig:classification}
\end{figure}

I also explored the performance of AnDi-ELM for the classification of trajectories among the 5 diffusion categories of the AnDi challenge, associated with the underlying diffusion model. An overview of the results is shown by the confusion matrix obtained for 1D, 2D, and 3D shown in Fig.~\ref{fig:classification}(a). AnDi-ELM correctly classifies a large fraction of trajectories, with a significant improvement of overall performance as the dimension is increased ($F_1$-score $=0.74$, $0.77$, and $0.85$ for 1D, 2D, and 3D, respectively). In particular, it showed excellent accuracy for detecting trajectories undergoing CTRW ($\sim 90$\%) and LW $ > 95$\%. Performance was also rather satisfactory for SBM, with a correct classification in $78-89$\% of the cases. The overall performance is largely affected by the false classification of FBM as SBM ($21-29$\%) and of ATTM as SBM ($\sim 38$\% in 1D and 2D). Interestingly, the capability for ATTM classification largely improves in 3D ($79$\%). The classification of FBM as SBM can be associated with the fact that both models have the same probability density function of displacements (Gaussian)~\cite{jeon2014scaled}. The identification of ATTM as SBM can instead be produced by the time-dependent diffusivity presented by both models. Moreover, for short trajectories not undergoing any change of diffusion coefficient, ATTM also shows a Gaussian probability density function.

As for the inference task, I investigated the dependence of the task metrics vs trajectory length and SNR (Fig.~\ref{fig:classification}(b)). Also in this case, as expected, I obtained a higher $F_1$-score for longer trajectories. However, the noise seems to play almost no role, only affecting 1D trajectories at low SNR.

Last, I calculated the $F_1$-score as a function of the trajectory length and the value of the anomalous diffusion exponent (Fig.~\ref{fig:classification}(c)). For this analysis, trajectories were pooled in three groups corresponding to strong subdiffusion ($\alpha<0.75$). Brownian and mild anomalous diffusion ($0.75 \leq \alpha < 1.25$), and strong superdiffusion  $\alpha \geq 1.25$). Except for very short trajectories, AnDi-ELM has a better capability of correctly classifying superdiffusion than subdiffusion. The performance further degrades for the classification of Brownian diffusion. However, it must be stressed that this result also reflects the different number of models associated with the diffusion modalities. In fact, only 3 models (FBM, LW, and SBM) are compatible with superdiffusion, whereas 4 (all except LW) can be subdiffusive, and all of them are compatible with Brownian diffusion.

\subsection{Trajectory segmentation}
Originally, I planned to develop AnDi-ELM only for the regression and classification tasks of the AnDi challenge. However, after the conclusion of the competition, I thought about applying the same method also for the segmentation task. This task had the objective of identifying the changepoint within a trajectory switching $\alpha$ and diffusion model, as well as determining the exponent and model for the identified segments. Here, I discuss the performance relative to changepoint identification. In fact, once the trajectory has been split, inference of the anomalous diffusion exponent and model classification can be performed with the methods described above.

For the detection of the changepoint (cp), I modified the feature engineering step from previous tasks, as described in Section~\ref{sec:methods}. The problem was tackled as a classification task. The AnDi-ELM provided scores for each timepoint, associated with the probability to host a changepoint. The timepoint presenting the maximum score along the trajectory is thus identified as a changepoint if it exceeds a given threshold.

\begin{figure}[h!]
    \centering
    \includegraphics[width=1\textwidth]{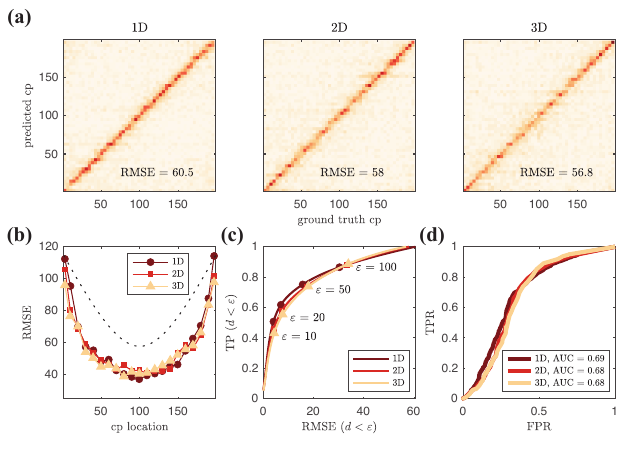}
    \caption{Results of the application of AnDi-ELM for the changepoint identification. {\bf (a)} Two-dimensional histograms of the frequency of changepoint predicted at a specific location as a function of the location of the ground-truth changepoint for trajectories of walkers undergoing anomalous diffusion in 1, 2, and 3 dimensions. {\bf (b)} Root mean squared error (RMSE) between the predicted and ground truth cp as a function of cp location in 1, 2, and 3 dimensions. The dashed line corresponds to random predictions. {\bf (c)} Plot of the fraction of true positive (TP) identifications vs the RMSE for prediction falling within a distance $\varepsilon$ from the true cp. {\bf (d)} Receiver operating characteristic (ROC) curves obtained for trajectories in 1, 2, and 3 dimensions. For the latter analysis, trajectories presenting a cp within 5 timepoints from both edges were considered as not undergoing any switch. Corresponding values of area under the curve (AUC) are reported in the legend.}
    \label{fig:segmentation}
\end{figure}

The two dimensional histograms shown in Fig.~\ref{fig:segmentation}(a) display a prevalence of identifications (predicted cp, $t_{\rm P}$ ) proximal to the true changepoint position (ground-truth cp, $t_{\rm GT}$). The RMSE was used to summarize the method's performance. As for the other tasks, the metrics show better results the higher the number of dimensions. As shown in Fig.~\ref{fig:segmentation}(b), the RMSE is not uniform with respect to the ground-truth cp location. This behavior is also observed for a random cp prediction (dashed line) and is associated with the breaking of symmetry produced by the cp. However, except for cp's located very close to the trajectories edges, AnDi-ELM shows a significant improvement with respect to random predictions.

To better investigate the performance of AnDi-ELM on cp detection, I performed further analyses. First, I studied the effect of applying a distance threshold $\varepsilon$ for the identification of a cp as a true positive. As $\varepsilon$ was progressively increased, only prediction falling within a distance $d = | t_{\rm P} - t_{\rm GT}| < \varepsilon$ from the ground truth  were taken into account for the calculation of both the fraction of correct identification (true positive, TP) and the RMSE (Fig.~\ref{fig:segmentation}(c)). Despite the expected correlation between the estimators, this analysis provides a practical indication for the choice of the distance threshold, depending on which estimator is more valuable for user applications. I also constructed receiver-operating characteristic (ROC) curves obtained at varying the score threshold for identifying the presence of a changepoint (Fig.~\ref{fig:segmentation}(d)). For this case, trajectories having a cp within 5 timepoints from the edge were considered as not having a cp. The AnDi-ELM shows the area under the curve (AUC) values $\sim 0.7$ and minor dependence on trajectory dimensions. 

\begin{figure}[!h]
    \centering
    \includegraphics[width=1\textwidth]{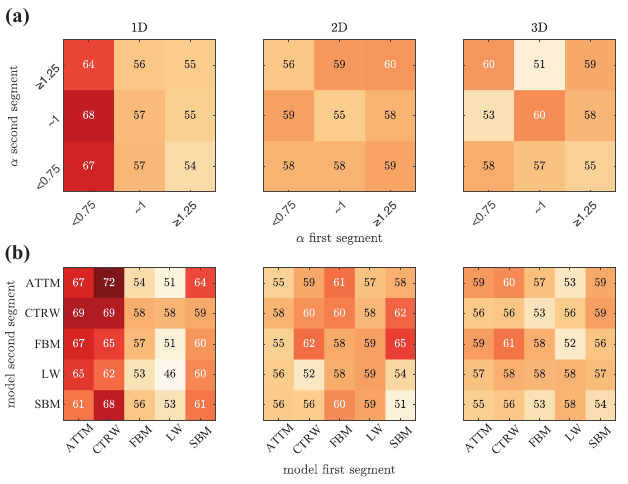}
    \caption{{\bf (a)} Root mean squared error (RMSE) between the predicted and ground truth cp as a function of the anomalous diffusion exponent of the first and second segment for trajectories in 1, 2, and 3 dimensions. {\bf (b)} Root mean squared error (RMSE) between the predicted and ground truth cp as a function of the diffusion model of the first and second segment  for trajectories in 1, 2, and 3 dimensions.}
    \label{fig:segmentation2}
\end{figure}

Last, I calculated the dependence of the RMSE as a function of the combination of anomalous exponents (Fig.~\ref{fig:segmentation2}(a)) and diffusion models (Fig.~\ref{fig:segmentation2}(b)) of the two segments composing the trajectories. Besides a slight degradation of performance observed in 1D when the first segment is strongly subdiffusive or undergoes ATTM or CTRW, no clear pattern emerges from these analyses.

\section{Conclusion}
I have described AnDi-ELM, a method for the quantitative characterization of anomalous diffusion from single trajectories according to the problems proposed in the AnDi challenge. The method applies a popular machine-learning approach to a set of features based on estimators from classical statistics. As such, it does not require previous information about trajectories. With minor modifications to the entry features, the method can be applied to the 3 tasks of the AnDi Challenge (\href{http://www.andi-challenge.org}{www.andi-challenge.org}), obtaining respectable results. AnDi-ELM ranked 11\textsuperscript{th} out of 13 teams in inference task in 1D, and 9\textsuperscript{th} out of 14 teams in the classification task in 1D. The extension of the method to provide predictions for higher dimensions and the segmentation task was implemented after the challenge submission deadline. Moreover, its performance is negligibly affected the magnitude of the noise. The main limitations of the method emerge in the inference of the anomalous diffusion exponent for ATTM and superdiffusive SBM; and in the discrimination of ATTM and FBM from SBM, mainly in 1D and 2D. Improvements in this sense might be achieved by applying a larger set of features. While most advanced machine learning techniques obtain better results, AnDi-ELM is straightforward to implement and extremely fast to train on a conventional laptop. Therefore, it might be the method of choice when a quick evaluation is needed, e.g., as a preliminary screening for the application of more advanced and time-consuming approaches.
\section*{Acknowledgments}
C.M. would like to thank Gorka Mu\~{n}oz-Gil for help with simulations; Pere Mart\'{i}-Puig and Jordi Sol\'{e}-Casals for introducing him to ELM; and the co-organizers of the AnDi challenge for stimulating discussions.
C.M. acknowledges funding from FEDER/Ministerio de Ciencia, Innovaci\'{o}n y Universidades -- Agencia Estatal de Investigaci\'{o}n through the ``Ram\'{o}n y Cajal'' program 2015 (Grant No. RYC-2015-17896), and the ``Programa Estatal de I+D+i Orientada a los Retos de la Sociedad'' (Grant No. BFU2017-85693-R); from the Generalitat de Catalunya (AGAUR Grant No. 2017SGR940). C.M. also acknowledges the support of NVIDIA Corporation with the donation of the Titan Xp GPU and funding from the PO FEDER of Catalonia 2014-2020 (project PECT Osona Transformaci\'{o} Social, Ref. 001-P-000382).

\section*{References}

\clearpage
\section*{Figures}
\end{document}